\documentclass[10pt,journal,compsoc]{IEEEtran}
\ifCLASSOPTIONcompsoc
  \usepackage[nocompress]{cite}
\else
  \usepackage{cite}
\fi
\ifCLASSINFOpdf
\else
\fi
\usepackage{listings}
\usepackage{amsmath}

\usepackage{graphicx}
\usepackage{caption}
\usepackage[caption=false]{subfig}
\usepackage{hyperref}

\newbox{\myorcidaffilbox}
\sbox{\myorcidaffilbox}{\large\includegraphics[height=1.7ex]{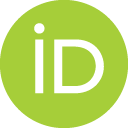}}
\newcommand{\orcidaffil}[1]{%
  \href{https://orcid.org/#1}{\usebox{\myorcidaffilbox}}}

\newcommand\numberthis{\addtocounter{equation}{1}\tag{\theequation}}
\begin{document}
%
\title{Energy hardware and workload aware job scheduling towards interconnected HPC environments}
%
%
%
%
\author{Marco~D'Amico~\orcidaffil{0000-0002-6195-6204},
         and~Julita~Corbal\'an~\orcidaffil{0000-0002-3926-5634}
\IEEEcompsocitemizethanks{\IEEEcompsocthanksitem M. D'Amico is with Barcelona Supercomputing Center (BSC), Barcelona, Spain. E-mail: marco.damico@bsc.es\protect\\
\IEEEcompsocthanksitem J. Corbal\'an is with Universitat Politecnica de Catalunya, Barcelona, Spain. E-mail: juli@ac.upc.edu}

\thanks{}}
\markboth{}%
{}
%



\IEEEtitleabstractindextext{%
\begin{abstract}

New HPC machines are getting close to the exascale. Power consumption for those machines has been increasing, and researchers are studying ways to reduce it.
A second trend is HPC machines' growing complexity, with increasing heterogeneous hardware components and different clusters architectures cooperating in the same machine. We refer to these environments with the term heterogeneous multi-cluster environments.

With the aim of optimizing performance and energy consumption in these environments, this paper proposes an Energy-Aware-Multi-Cluster (EAMC) job scheduling policy. EAMC-policy is able to optimize the scheduling and placement of jobs by predicting performance and energy consumption of arriving jobs for different hardware architectures and processor frequencies, reducing workload's energy consumption, makespan, and response time. The policy assigns a different priority to each job-resource combination so that the most efficient ones are favored, while less efficient ones are still considered on a variable degree, reducing response time and increasing cluster utilization.

We implemented EAMC-policy in Slurm, and we evaluated a scenario in which two CPU clusters collaborate in the same machine. Simulations of workloads running applications modeled from real-world show a reduction of response time and makespan by up to 25\% and 6\% while saving up to 20\% of total energy consumed when compared to policies minimizing runtime, and by 49\%, 26\%, and 6\% compared to policies minimizing energy.
\end{abstract}

}

\maketitle

\IEEEdisplaynontitleabstractindextext

%
\IEEEpeerreviewmaketitle

\IEEEraisesectionheading{\section{Introduction}\label{sec:introduction}}

\IEEEPARstart{O}{ver} the last years, we have witnessed an explosion of hardware heterogeneity. Starting from GPUs used for general purpose, then a high number of processors thanks to instruction sets like ARM~\cite{arm}, Power ISA~\cite{powerisa}, RISC-V~\cite{riscv}, and finally special-purpose architectures, e.g., Tensor Processing Units (TPUs) are contributing to increasing the heterogeneity and complexity of HPC machines. 

In HPC, cited architectures collaborate in computing nodes, which in turn are grouped in clusters. In some cases, different clusters cooperate in the same machine. We call those environments \textit{heterogeneous multi-cluster environments}. A practical example is an old machine still active, like in the case of Marenostrum III\cite{mn3} and Minotauro\cite{mt}, or heterogeneous clusters like Power9, running aside Marenostrum IV\cite{mn4}. While these machines have their own resource manager, the author hypothesizes that interconnecting the clusters under one resource manager can greatly improve overall utilization. 

Another example is the prototype machine developed in DEEP-EST project~\cite{deep}, proposing a number of heterogeneous clusters collaborating under the same machine and managed by a unique resource manager. Each cluster has general-purpose CPUs and accelerators, presenting different performances and energy behaviors.

From a broader perspective, with the advance of memories and interconnections, multiple machines will likely be connected and exchanging workloads, resulting in machines with huge potential but also high hardware diversity and more complex management of resources.

On the other side, HPC machines are extremely power demanding, such that the US. Department of Energy set a goal of 20 MW power consumption for an exascale machine. In recent years research and industry put a high effort investigating how to stay under this constraint. At the same time, there is a significant shift from classical performance-oriented to an efficiency-oriented research mentality, with increasing awareness of green and eco-friendly concepts.
Top500~\cite{top500} list now is accompanied by the Green500~\cite{green500}, classifying most energy-efficient machines. New metrics are proposed to draw up the list, i.e., TGI metric~\cite{6270748} as an alternative to GFLOPS/W, which evaluates non-computing devices, e.g., memories and disks.

Research in energy consumption moved in both hardware and software directions.\\
Regarding hardware techniques, miniaturization plays a vital role in reducing power consumption, but this process is getting more and more to the silicon physical limit, and alternative chemical elements are being studied. Meanwhile, Dynamic Frequency and Voltage Scaling (DVFS) became a popular solution to reduce consumption by dynamically modifying the processor's voltage and frequency. This technique is beneficial when trying to control the maximum power consumption or when dealing with memory-bounded applications. In the second case, the so-called memory wall, related to main memory limited bandwidth and high latency, prevents higher frequencies from benefiting the runtime.

Regarding software techniques, we can classify them in the following categories, from a low-level to a high-level perspective of the system:
\begin{itemize}
    \item Low level and OS interfaces for DVFS, collection of power metrics.
    \item Energy modeling and estimation for hardware and applications.
    \item Thermal, power, and energy-aware task scheduling in multi-core processors and GPUs, node-level powercapping.
    \item Thermal, power, and energy-aware job scheduling and resource management at cluster level: two main techniques are used to limit power consumption: (1) Overprovisioning: it is a strategy in which more hardware than the one that can be powered is bought, and part of it is selectively shut down or powered on based on the necessity. (2)Powercapping at the node, job, and system level.
\end{itemize}

We identified that most of the research~\cite{8425478} is mainly focusing on power scheduling and controlling, with few considerations on the fact that minimizing power does not mean minimizing the energy. While power-awareness is essential, the authors' opinion is that power-saving solutions are related to a constraint in HPC machines' design. On the other side, an energy-aware scheduler needs to optimize the energy-performance trade-off continually. For this reason, power-awareness and energy-awareness techniques are not exclusive, and they can be efficiently combined.

Furthermore, as suggested by recent research~\cite{Czarnul2019EnergyAwareHC}, this whole variety of solutions miss automatic ways of configuring the systems and user's activity. In particular, the job scheduling and resource management layer manages all the resources and gives an interface to users to use them. At the current state of the art, we identified that it is the user's responsibility to specify the type of resource needed, the number of resources, and the best energy settings for the submitted job. This complexity is not acceptable in heterogeneous multi-cluster environments, where several computing node types coexist, each one with different characteristics, resulting in confusion and unnecessary knowledge needed for users.

Filling those gaps, we propose the Energy-Aware Multi-Cluster scheduling policy (EAMC-policy). Enabling per-job energy-performance characterization and comparison on heterogeneous environments at job scheduling level, EAMC automatizes jobs' placement and selection of optimal clock frequencies, respecting a trade-off between performance, energy consumption, and response time. 
For this purpose, we extended an energy model from the related work\cite{ear_model} to characterize applications' runtime and energy on different hardware resources. Based on the extended model, we implemented a multi-objective energy and performance classification. We used it to assign a different priority to each job-resource combination, favoring the optimal ones but keeping the non-optimal to balance the load and reduce response time.
EAMC-policy reduces user's overhead and required expertise, error-prone information, and eventual malicious behavior by automatizing the process of selecting frequency and optimal hardware.

We integrated it into Slurm\cite{10.1007/10968987_3}, a Distributed Resource Management System (DRMS), and the Slurm Simulator\cite{8641556}.\\
Integrating the prediction model in a job scheduling simulator, we made the Slurm Simulator workload and energy-aware, capable of calculating energy consumption based on the type of application and not only the hardware like most of the job scheduling simulators. We modeled various applications and benchmarks behavior in terms of energy and performance for multiple architectures, and we distributed them in a workload generated with Cirne's model\cite{cirne}.

We studied the case of a heterogeneous two-cluster environment, each one equipped with processors with different characteristics, i.e., number of cores, frequencies, and cache size. We evaluated performance by changing the distribution of jobs that favor the first and the second cluster. By running job scheduling simulations, we observed improvements in energy consumption by up to 20\% while reducing response time by 25\% and makespan by 6\%.

This document is structured as follows: Section~\ref{RW} resumes the state of the art of the energy-aware job scheduling topic, Section~\ref{en_modeling} describes the preliminary work on job modeling, energy tools, and interfaces. Section~\ref{ea_policy} explains the EAMC-policy in detail, while Section~\ref{evaluation} evaluates it and compares variants of the policy and its parameters with the standard version of the DRMS. Finally, Section~\ref{conclusion} resume and conclude the document, with insights on the future work.

\section{Related Work}\label{RW}
Several surveys exist and resume the work done in the energy field for HPC. Czarnul et al.~\cite{Czarnul2019EnergyAwareHC} give an overview of leading energy-aware HPC technologies, diversifying computing environments, device types, metrics, benchmarks, energy-saving methodology, and energy simulators.

Regarding energy-aware job scheduling, Maiterth~\cite{8425478} resumes some of the leading HPC centers' energy strategies and their future directions. Most centers work on powercap solutions, a few are working on overprovisioning, and two investigate energy-aware solutions. We found this is a general tendency in the research, with powercap adopted as the primary strategy. In our vision, optimizing energy consumption is crucial, while power limits are a constraint due to machines' powering.

We identified the scarcity of research on energy-aware job scheduling solutions for heterogeneous multi-cluster environments. The authors consider this is a vital field to investigate, given the potential and complexity of those environments.

Netti et al.~\cite{10.1007/978-3-319-92040-5_1} explore the effect of different ways of prioritizing critical resources, the scarcest and most demanded. While this work brings improvement for heterogeneous clusters, it is not energy and workload aware. Extending this work to include energy-aware prioritization could lead to interesting results.

Some policies are based on a model that is aware of manufacturing and assembling variability. Chasapis et al.~\cite{10.1145/3330345.3330372} propose different scheduling policies considering processor manufacturing variability under a power constraint. Moore et al.~\cite{10.5555/1247360.1247365} propose a temperature-aware workload placement by detecting cooling inefficiencies that may come from places relatively distant from the temperature sensor, and it tries to minimize heat re-circulations. EAMC can use per-node energy models to model the manufacturing variability.

Sarood et al.~\cite{7013053} combine malleability and DVFS to create a scheduling policy that adapts the workload to a strict power budget in over-provisioned systems. Similarly, in a precedent research~\cite{10.1145/3229710.3229752, 10.1145/3337821.3337909}, we used malleability and node sharing techniques to reduce response time, makespan, and energy consumption. The integration of the policies is a promising research direction.

Barry et al.~\cite{10.1145/1088149.1088179} explore overprovisioning by powering down nodes in a controlled way using online simulation and controlling system slowdown, keeping it to acceptable values. This method works well on non highly loaded systems, but it is difficult to exploit in the opposite case.

Borghesi et al.~\cite{BORGHESI20181} implement hybrid scheduling techniques for systems under a power cap. They perform a power estimation by using machine learning techniques on historical data instead of a per-job characterization. While this interesting approach does not require the user to specify energy tags, as required by our model, it is not easy to obtain the same per-job precision as our approach.

In the research of Rajagopal et al.~\cite{8287744}, based on this work~\cite{10.1145/2749246.2749262}, researchers integrate into Slurm~\cite{10.1007/10968987_3} a power-aware scheduler that allows setting powercaps and collects power metrics at runtime to adjust them. It incorporates an interface to choose a frequency based on user hints and powercap constraints. The solution works with a basic power model using hardware information and user guidance, with estimations representing an upper limit and not the actual power consumption. Our research differs since it optimizes energy, not power. Moreover, no energy input from users is required to set the best frequency, but it is predicted and set automatically using online metrics collection and historical data. In their work, DVFS is used when reaching the powercap, while in our work, we use DVFS to select the optimal performance-efficiency trade-off for each job. In future work, the two policies could be combined, i.e., by choosing optimal frequency for performance and efficiency while assuring a powercap.

Auweter et al. implemented two policies: energy-to-solution and best-runtime, to select the most suitable frequency for jobs~\cite{10.1007/978-3-319-07518-1_25}. Users are required to specify tags for similar jobs to be able to use the proper job characterization. This work is the base of our research, as we implemented a similar description of runtime, energy, and power for jobs. On our side, our instrumentation allows us to verify the correct use of energy tags and recalculate optimal frequencies at runtime. We implemented it in Slurm. We extended the energy model to characterize jobs on heterogeneous clusters. We propose the EAMC-policy to prioritize the most efficient architectures following an energy-performance trade-off strategy instead of energy-to-solution or best-runtime.


\section{Preliminary work}\label{en_modeling}
EAMC-policy selects optimal energy configuration and places jobs based on energy and performance estimations. 
EAMC requires estimating the job's runtime and power consumption at scheduling time before the job start. In this section, we describe the designed framework that allows EAMC-policy operation.

\subsection{Job power and runtime modeling}
To estimate runtime, power, and energy consumption for
different processors, frequencies, and applications, we revised available energy models from the state of the art.
Various energy models and tools for energy prediction are proposed in the literature, with distinct complexity.
Some of them are general, based on hardware information only, whereas the chosen~\cite{ear_model} is able to model different applications' behavior for different architectures. The model needs two inputs:
\begin{enumerate}
    \item application data: a collection of metrics that characterize applications. The model uses runtime, average consumed power, average number of memory transactions per instructions (\textit{TPI}), and average cycles per instruction (\textit{CPI}) at the reference frequency f\textsubscript{ref}.
    \item hardware coefficients: a set of parameters obtained by running a learning phase based on a set of benchmarks. The chosen model uses six coefficients learned using linear regression: \textit{A, B, C, D, E, F}.
\end{enumerate}
Equations~\ref{eq:power},~\ref{eq:cpi}, and ~\ref{eq:time} describe the mathematical model that estimates power consumption \textit{P} and runtime \textit{T} at the frequency \textit{f}, starting from a default frequency f\textsubscript{ref} at which application data is given.

\begin{align*}
P(f)= A * P(f_{ref}) + B * TPI(f_{ref}) + C
\numberthis
\label{eq:power}
\end{align*}

\begin{align*}
CPI(f) = D * CPI(f_{ref}) + E * TPI(f_{ref}) + F
\numberthis
\label{eq:cpi}
\end{align*}

\begin{align*}
T(f) = T(f_{ref}) * \frac{CPI(f)}{CPI(f_{ref})} * \frac{f_{ref}}{f}
\numberthis
\label{eq:time}
\end{align*}

To implement and use the energy model in a real-world environment, we used Energy-Aware Runtime (EAR)~\cite{ear}.\\
EAR is a framework that provides an energy-efficient solution for HPC clusters. It includes monitoring, accounting of the applications' performance, and it integrates energy optimization via DVFS at node and cluster level. It is deployed on SuperMUC-NG~\cite{supermuc} and implements the previously described energy model.

EAR was used to learn hardware coefficients \textit{A, B, C, D, E, and F} to model two computing nodes equipped with general-purpose processors with a different number of cores and frequencies, as described in Section~\ref{evaluation}.

In a non-simulated environment, EAR collects application data dynamically at runtime and stores them in the application database to be used by the energy model. In our test simulated environment, we use static data collected by running several applications with diversified computing and memory behaviors in a previous phase. Used applications are described in Table\ref{tab:apps}.

To evaluate the proposed policy in our test system, we used and extended an interface developed in the context of the DEEP-EST project~\cite{e-interface}.

The interface is configurable with different machine configurations and energy models. DRMS uses it to retrieve energy, time, and power estimations. It can be used in a real environment as a standard interface that abstracts underlying energy models or to simulate tools like EAR in simulated environments.

To run our simulated environment, we implemented the chosen prediction model by extending the interface. Each hardware configuration is identified by a model id, a series of hardware coefficients, a node configuration, and a range of frequencies. Different hardware configurations can exist for the same node, e.g., a node using or not using the accelerator or a processor using or not using the vector extension.

\subsection{Application Database}
The Application Database \textit{AppDB} stores application metrics collected at runtime for each hardware configuration.
For each tuple, appDB contains fields as the appID, the modelID, the necessary application metrics.

In a real system, application database entries for jobs are created on the fly. In the first run, there is no energy information associated with the job. Still, metrics are collected by EAR, which creates an entry in the DB associated with the user-specified appID. For the first run, optimal frequency is not estimated previously, but it is set at runtime as soon as data is available.

In the case of multiple clusters but with application data related to only one of the clusters, performance and energy consumption can be estimated from data in the appDB in combination with the hardware model, or they can be evaluated when the application runs for the first time on the hardware.

In our simulated environment, for the sake of simplicity, the appDB is static and available at the beginning of simulations. To obtain the application's data, we run applications in different architectures, collect necessary metrics using the EAR accounting and store them in the database.


\subsection{Model extension for multi-cluster environments}
When collecting metrics, while hardware counters and average power are automatically collected by EAR, the application's runtime is potentially variable in each run.

From the DRMS point of view, it is not straightforward to estimate the runtime parameter with precision, so commonly, users are asked to give a requested time for the job. These values are usually far from the real job's runtime, and it can also be a default value in the case the user does not specify it.

Having requested time as the only time-related information for the job, the energy model assumes this value relates to the primary partition. To calculate a requested time for the other partitions, we extended the energy model to learn a seventh parameter, \textit{time\_coefficient}. time\_coefficient represents the ratio between the main and another partition runtime. This parameter is estimated by running the same application on both hardware architectures. Future work will still consider more complex models that do not need this phase, e.g., based on already collected hardware metrics.

\subsection{User interface for job submission}
When submitting a job, users can avoid specifying which hardware architecture they want to use if they accept that the job will run on nodes automatically selected by the scheduler.

We evaluated EAMC on processors from the same family, so we automatically consider the app can run on every available hardware architecture, thanks to binary compatibility.
Considering similar architectures is not far from the case of more heterogeneous ones, given the high effort of research and industry in supporting interoperability, hardware-independent programming languages, and libraries. In the case there is no binary compatibility, users can specify different binaries for the same appID. This information can be stored in the appDB or contained in the job script in a format readable by a job script parser.

Users are still in charge of specifying the number of requested nodes and processes. The number of threads or processes per node can be configured by using Slurm plugins or precedent work~\cite{10.1145/3229710.3229752, 10.1145/3337821.3337909} more intelligently. If a memory requirement is specified and some nodes cannot satisfy this requirement, a new number of nodes will be calculated.
 
Using more heterogeneous architectures, on the other side, increases the complexity of parameters for the job. For example, the number of requested nodes needs to increase to maintain a reasonable runtime when switching from a fast and power-consuming resource to a slow and less demanding one. Future work will try to take into account those details by increasing the complexity of the hardware models.

It is also the user's responsibility to specify an \textit{appID} in the job script to recognize jobs with the same behavior. On the other side, EAR can verify appID related history corresponds once the job starts, giving the DRMS instruments to detect and penalize fraudulent users' behavior. Suppose runtime collected metrics do not conform to stored information in the appDB for the specified appID. In that case, the application keeps running, and a new frequency is calculated based on runtime data. If the user repeatedly uses wrong appIDs, the DRMS can detect it by analyzing his history of jobs, return a warning message, or take additional actions.

\section{Energy-aware job scheduling in multi-cluster environments}\label{ea_policy}
EAMC-policy is based on priority backfill, with energy predictions influencing the priority of arriving jobs.
\footnote{EAMC code is publicly available~\cite{slurmsimcode}}. This section describes the implementation of the policy.

In the policy presentation, for simplicity, we use the term \textit{partitions} as a logical representation of resources inside the DRMS. While ideated with the introduced concept of multi-clusters, the policy is generic, independent of the physical and logical representation of the hardware, either physically or non physically separated, grouped, or non grouped in partitions.

\subsection{Problem definition}
We model the proposed policy as an online job scheduling problem. While this problem usually has as the objective function the minimization of the makespan, in our case, we describe it as a multi-objective optimization problem with objectives of the reduction of makespan and energy consumption.

The most efficient solver for job scheduling problems in the literature is the backfill algorithm. We used a backfill version with priorities, where each job gets assigned a priority based on several factors, e.g., arrival time, requested number of nodes.

To make backfill energy aware, we included a runtime and energy classification of jobs as a further factor for priorities, thus influencing the scheduling order. More precisely, we calculate a priority for each \textit{job-partition} combination \textit{jp}, defined as a job \textit{j} running on a partition \textit{p} in two phases using Equation~\ref{eq:bestfreq} and~\ref{eq:bestpart}. The first equation is used to estimate the optimal processor frequency \textit{f\textsubscript{opt}}, while the second is used to classify and assign a different priority to each p for j.

Firstly, in Equation~\ref{eq:bestfreq}, given a \textit{jp}, with the objective of calculating the optimal frequency \textit{f\textsubscript{opt}}, we define the metric \textit{dist\textsubscript{jp}} as the Euclidean distance from the origin in the Euclidean space representing time and energy normalized to the minimum value of the respective metrics (E\textsuperscript{min}\textsubscript{jp}, t\textsuperscript{min}\textsubscript{jp}) in the prediction space of \textit{jp}. \textit{t\_weight} parameter increases or decreases runtime estimation weight over energy.

\begin{align*}
dist_{jp} = min \sqrt{\frac{E_{jp}(f)}{E^{min}_{jp}}^2 + t\_weight * \frac{t_{jp}(f)}{t^{min}_{jp}}^2},\\
\forall f \in [f_{min},f_{max}]\numberthis \label{eq:bestfreq}
\end{align*}

As a second point, in Equation~\ref{eq:bestpart}, given multiple job-partition entries for a job, to classify partitions, we used the Euclidean distance again to calculate \textit{part\_eff}, in this case normalizing energy and time among all partitions of the job \textit{j} (E\textsuperscript{min}\textsubscript{j}, t\textsuperscript{min}\textsubscript{j}) at frequency f\textsubscript{opt} calculated for each partition with Equation~\ref{eq:bestfreq}. As a result, \textit{part\_eff} can be used as an indicator of performance and efficiency, and it can be used to influence the \textit{jp} priority. Again, \textit{t\_weight} affects the weight of time over energy.
\begin{align*}
part\_eff = \sqrt{\frac{E_p(f_{opt})}{E^{min}_{j}}^2 + t\_weight * \frac{t_p(f_{opt})}{t^{min}_{j}}^2},\\
\forall p \in partitions\numberthis \label{eq:bestpart}
\end{align*}

\subsection{Proposed policy}
Figure~\ref{fig:reorder} and~\ref{fig:priorityinc} represent two variants of the EAMC-policy.
First, a job, represented on the top left, is submitted. At submission time, the policy assigns the priority to the arriving jobs and, in the case of multiple partitions, to each job-partition combination. We extended this part of the DRMS with the Energy-prediction Priority Module (EPM).
Once a different priority for each job-resource is assigned, an EAMC-Scheduler (EAMCS) schedules jobs, from a unique priority queue, in order of priorities. Following on, we describe EPM and EAMCS components.

\subsubsection{EAMC-Scheduler (EAMCS) and Energy-prediction Priority Module (EPM)}
EAMCS, based on EASY-backfill~\cite{932708}, tries to schedule one job-partition per time in order of priority. Using backfill allows scheduling jobs considering a trade-off between system energy, performance, and average wait time. For example, if the higher priority job cannot run due to lack of resources or a big enough time window for backfilling it, the lower priority job-partition can start if enough resources are available. While not being the best solution in terms of energy and runtime, it is optimal when considering the wait time. Finally, following the Slurm implementation, during backfill, holes in the scheduling reserved by high priority jobs cannot be used by lower priority jobs, and job-partitions from the same job do not influence each other as they request different resources.

The proper equilibrium of jobs running on the favored and non-favored partition can lead to optimal performance in the time-energy dimension. Extreme behaviors, where jobs poorly performing are discarded, are managed in one of the policy alternatives described in this section, with details on EAMCS.

Listing~\ref{alg:energyplugin} describe EPM implementation in detail. In the first inner loop, EPM gets runtime and energy predictions using the energy interface and energy models (line 5). The parameter \textit{time\_coefficient} is used as a multiplier to predict runtime and energy consumption for different partitions. Then, using Equation~\ref{eq:bestfreq}, it individuates the best frequency for each job-partition, setting it as the default value (line 6). The same loop calculates \textit{E\textsuperscript{min}\textsubscript{j}} and \textit{t\textsuperscript{min}\textsubscript{j}}.

Once frequencies are picked, the second loop evaluates and sorts partitions according to Equation~\ref{eq:bestpart} (lines 11-13). In the last loop, \textit{set\_priority()} function (line 16) is in charge of assigning a priority to each partition according to the selected strategy. In the end, the job-partition couple enters the job priority queue.

\begin{lstlisting}[basicstyle=\small,language=Python, caption={Energy Prediction Module (EPM) implementation}, label={alg:energyplugin}, float, floatplacement=H, numbers=right, tabsize=2, breaklines=true]
for each j in submitted_jobs:
    if (!j->partitions)
        j->partitions = get_partitions(
            j->app_id);
    for each p in j->partitions:
        EA-API_get_predictions(p,j->app_id);
        p->best_f_val = get_optimal_freq(p,
            p->energy, p->time);
        if (p->energy < emin)
            emin = p->energy;
        if (p->time < tmin)
            tmin = p->time;
    for each p in j->partitions:
        p->part_eff = get_part_efficiency(
            p->energy,p->time,emin,tmin);
    sort(j->partitions,cmp_part_eff());
    
    for each p in j->partitions:
        p->priority = set_priority(p,policy_type);
        enqueue(job_queue,j);
\end{lstlisting}

\begin{figure}
    \centering
        \includegraphics[scale=1]{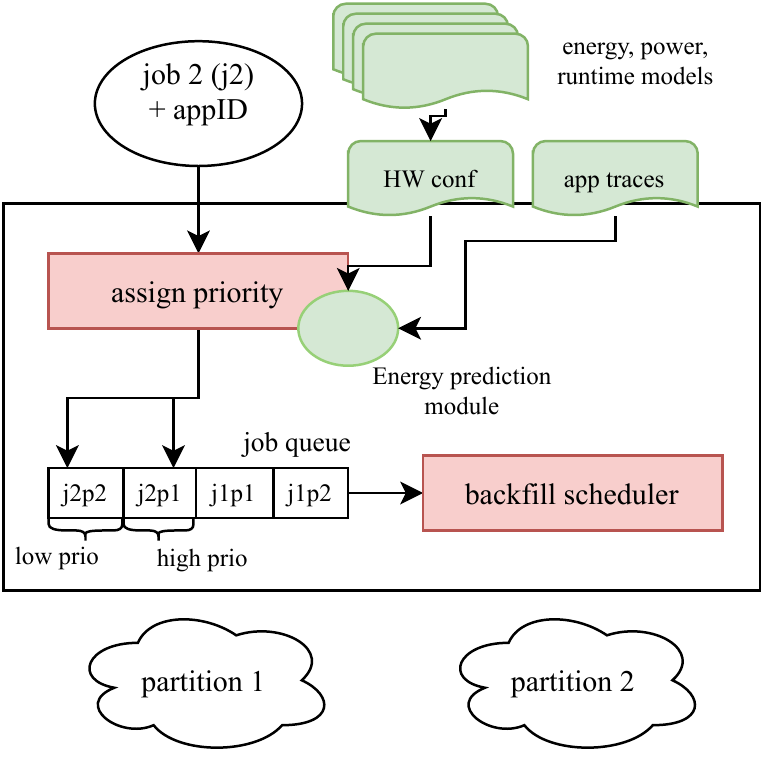}
    \caption{Job submission and scheduling phases for EAMC-Reorder policy. The big-box represents the DRMS. Red boxes are modified parts, and green boxes are energy prediction components. With EAMC-Reorder, jobs-partition entries are reordered while keeping the same priority in the queue.}
    \label{fig:reorder}
\end{figure}

\begin{figure}
    \centering
        \includegraphics[scale=1]{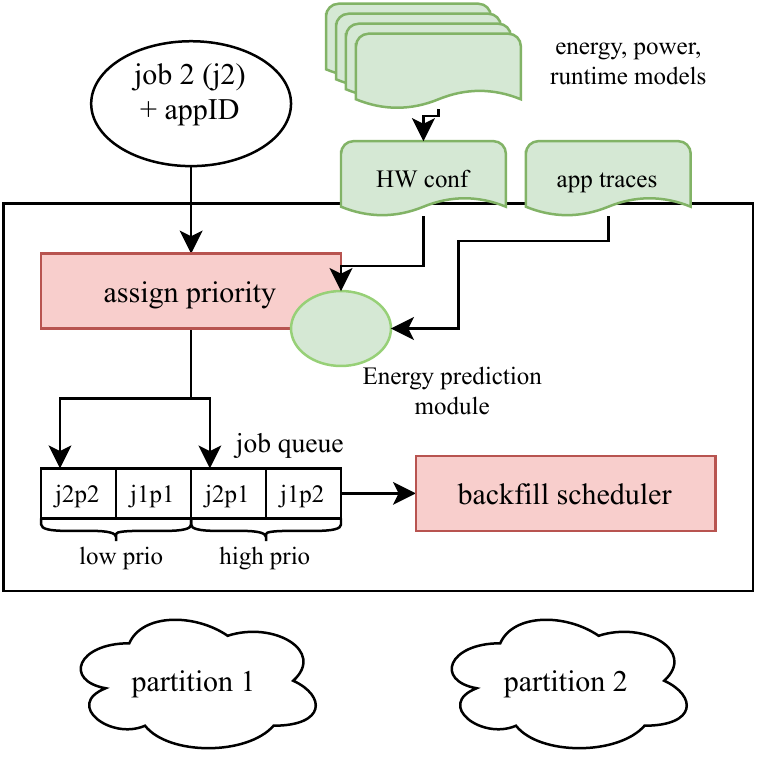}
    \caption{Job submission and scheduling phases for EAMC-PriorityInc policies. The big-box represents the DRMS. Red boxes are modified parts, and green boxes are energy prediction components. With EAMC-PriorityInc and EAMC-PriorityInc-V, the job queue is separated into higher priority and lower priority queues depending on energy and performance evaluations.}
    \label{fig:priorityinc}
\end{figure}

We developed three strategies that implement different ways of assigning priorities to job-partitions, depending on how aggressive is the policy in favoring optimal partitions over arrival order of jobs:
\begin{itemize}\nobreak
    \item \textbf{EAMC-Reorder}: it changes the order of partitions for \textit{the same job}, prioritizing job-partition with better performance. As a consequence, EAMCS maintains the arrival order of jobs while reordering partitions within the job. Jobs-partition entries are queued similarly to job2 in Figure~\ref{fig:reorder}.
    \item \textbf{EAMC-PriorityInc}: it assigns a different priority to each job-partition based on the position in the sorted partition list. As shown in Figure~\ref{fig:priorityinc}, the two job-partition are inserted in different points of the job queue, creating two separate queues, one with preferred partitions, the others for non-optimal partitions. In this case, EAMCS favors optimal job-partitions over non-optimal ones, giving less weight to the arrival time order.
    \item \textbf{EAMC-PriorityInc-V}: similarly to PriorityInc, based on the same figure, it assigns a different priority to each job-partition, based on the sorted partition list, with the difference that the priority is lowered if the evaluated metric is below a certain threshold. We set the threshold to 35\% from the optimal part\_eff metric. As an example, if \textit{j1p1} exceeds the threshold and \textit{j2p2} does not, \textit{j1p1} would go at the bottom of the queue. EAMCS first checks all optimal job-partition entries, then the second choices, and finally all the job-partitions below the threshold.
\end{itemize}

\section{Evaluation}\label{evaluation}
The evaluation is based on simulations using the BSC Slurm jobs scheduler Simulator~\cite{8641556}. We modeled a workload of 5000 jobs, with a makespan between 10 and 15 days, depending on applications distribution and used frequency, using Cirne model~\cite{cirne} configured with ANL arrival pattern. Jobs' average requested nodes are 18.29, and jobs' average duration is 3.65 hours at the highest frequency.
We run the workload in the following simulated clusters:
\begin{enumerate}
    \item p1: 512 nodes equipped with: 2x Platinum 8168 CPU [2.70GHz-1.20GHz] 24C, TDP 205W and 12 x 16GB DDR4 SDRAM
    \item p2: 512 nodes equipped with: 2x Gold 6254 CPU [3.10GHz-1.20GHz] 18C, TDP 200W and 12 x 32GB DDR4 SDRAM
    \item p3: 512 nodes equipped with: 2x Gold 6148 [2.4GHz-1GHz] 20C, TDP 150W and 12 x 16GB DDR4 SDRAM
\end{enumerate}
The number of modeled hardware architectures was limited by the available architectures and permissions needed to collect the necessary data. We sized the number of simulated jobs and system size according to the time limits imposed by the testing environment.

We run the EAR learning phase to get coefficients for the simulated nodes on Lenox\cite{lenox} cluster.
We used data from eight applications, running them on one node using all the available cores~\footnote{Energy models and applications data is publicly available~\cite{e02v-y602-20}}. Table~\ref{tab:apps} describes the set of applications, presenting different memory and compute profiles, reported through average cycles per instruction (CPI), memory bandwidth (GB/s), the ratio of the runtime and power between partitions, and the ratio between the delta between the maximum and minimum runtime and the delta of frequency $\Delta$r/$\Delta$f. The higher the value of $\Delta$r/$\Delta$f, the more the runtime scales with the frequency, e.g., ep.D runtime is highly affected by changes in the processor's frequency, while STREAM is not. 

We simulated two different systems:
\begin{enumerate}
    \item A system made up of p1 and p2.
    \item A system made up of p1 and p3.
\end{enumerate}
When comparing p1 and p2, seven out of eight applications prefer the first partition in terms of performance. In terms of power, the first partition showed slightly higher power consumption for the CPU component, more than the nominal 5 Watts, while the second partition showed up to double DRAM consumption compared to the first. From an energy perspective, applying Equations~\ref{eq:bestfreq} and~\ref{eq:bestpart}, for the tested t\_weight values, the first seven apps favor the first partition, while app 8 favors the second. For this comparison, app8, STREAM, is a weak scaling benchmark, where an amount of memory is allocated per-process, i.e., per-core. The second partition, having a reduced number of cores, has fewer data to manage, explaining the difference in runtime while showing similar hardware metrics. This is the typical behavior of some memory-bounded applications. EAMC will be aware of STREAM behavior and prioritize the partition with a lower number of cores.

For the same comparison, we distributed the applications among the workload by randomly drawing samples from three distributions:
\begin{enumerate}
    \item 13\% benefits from partition 2: while the distribution is uniform among all applications, 87\% of the workload favors the first partition.
    \item 33\% benefits from partition 2: In this 33\% of jobs run app8, which favor the second partition, the remaining 67\%, uniformly distributed among remaining apps, prefer the first.
    \item 50\% benefits from partition 2: finally, this case evaluates an even load among the two partitions in terms of the number of jobs per favored resource. 50\% of jobs run app8.
\end{enumerate}

Comparing p1 and p3, at base frequency, five out of eight applications run optimally on partition 1, while at the optimal frequency, all the applications run optimally on the same partition. In this case, we used a strong scaling version of STREAM that uses a fixed amount of memory per computing node. For this comparison, we run the Workload 13\%.

\begin{table*}
\centering
\begin{tabular}{cccccccc}
\textbf{AppID} & \textbf{Name} & \textbf{CPI} & \textbf{GB/s} & \textbf{Runtime (s)} & \textbf{Power (W)} & \textbf{$\Delta$r/$\Delta$f ($s^2$)} \\ \hline 
1 & lu.C    & 0.67 / 0.66 / 0.57 & 70.95 / 63.73 / 59.7  & 52.49 / 57.45 / 62.78 & 386 / 367 / 331.25  & 2.38 / 2.87 / 4.18 \\ 
2 & ep.D    & 0.60 / 0.60 / 0.60  & 0.03 / 0.04 / 0.03   & 64.20 / 74.20 / 86.59 & 349 / 337 / 290.32  & 5.67 / 6.37 / 8.97 \\ 
3 & bt-mz.C & 0.38 / 0.39 / 0.38  & 23.33 / 19.96 / 17.7  & 43.95 / 51.32 / 57.70 & 417 / 411 / 342.73 & 3.52 / 4.05 / 5.65 \\ 
4 & sp-mz.C & 0.44 / 0.50 / 0.42  & 66.09 / 52.24 / 51.17 & 46.05 / 59.76 / 58.58 & 460 / 427 / 375.47 & 3.07 / 3.34 / 4.64 \\ 
5 & lu-mz.C & 0.62 / 0.62 / 0.63  & 19.87 / 21.14 / 18.89  & 59.21 / 61.41 / 65.98 & 355 / 360 / 308.81 & 4.75 / 4.70 / 6.33 \\ 
6 & ua.C    & 0.99 / 0.95 / 0.87  & 51.29 / 49.05 / 45.5 & 46.96 / 53.11 / 54.51 & 368 / 360 / 324.6 & 1.74 / 1.84 / 2.64 \\ 
7 & DGEMM   & 0.40 / 0.40 / 0.38  & 66.43 / 62.76 / 50.87 & 47.40 / 52.60 / 61.97 & 491 / 497 / 366.25  & 3.29 / 3.08 / 4.09 \\ 
8 & STREAM  & 4.46 / 3.67 / 3.55  & 138.87 / 137.47 / 136.86 & 107.55 / 81.43 / 110.34 & 381 / 365 / 330.52 & 0.16 / 0.06 / 0.13 \\ \hline 
\end{tabular}%
\caption{Set of applications and their characteristics. Metrics specified in the order of partitions, in the format p1 / p2 / p3.}
\label{tab:apps}
\end{table*}
The evaluation follows in this section, where we analyze the performance of the developed policies compared to \textit{Base Slurm}, and policies inspired by Auweter~\cite{10.1007/978-3-319-07518-1_25}, called \textit{Min\_energy} and \textit{Min\_runtime}.

Base Slurm is configured to run jobs at default frequency, i.e., the maximum frequency, and each job is submitted only to the optimal partition.

Min\_runtime runs jobs at the frequency that minimizes the runtime, while Min\_energy runs jobs at the frequency minimizing the job's consumed energy. For those policies, jobs are submitted to both partitions, and each job-partition gets the same priority. The scheduler tries to run the job in the default order, first in the first specified partition and immediately after in the second, not establishing the favored one.

Evaluated metrics:
\begin{itemize}
    \item Makespan: The difference between last job end time and first job arrival time.
    \item Average response time: The average of jobs' response time, defined as the difference between end time and arrival time
    \item Sum of the jobs' consumed energy: sum of individual jobs' consumed energy, not including idle nodes.
    \item Sum of apps runtime: the sum of individual jobs' runtime, used to understand the impact of EAMC-policies on the jobs' performance.
    \item Avg frequency: the average of picked frequencies for the whole workload, or per partition.
    \item Percentage of apps in the optimal partition: The percentage of jobs scheduled in the favorite partition.
\end{itemize}

We first analyze the system p1/p2. We compare the EAMC policies variants to Base Slurm, Min\_runtime, and Min\_energy. We then compare Min\_runtime and Min\_energy to our policy when changing the applications' distribution in the workload. Finally, for system p1/p3, we analyze the 13\% case. For both systems, p1/p2 and p1/p3, we analyze the performance when changing the t\_weight parameter, i.e., giving equal or 1.5x more importance to performance over energy efficiency. 

\subsection{Comparing to Base Slurm}
This section compares EAMC-policy variants' performance with Base Slurm, where jobs are submitted only to the optimal partition, and with Min\_runtime and Min\_energy, where jobs are submitted to multiple partitions with a frequency that minimizes runtime or energy. In Base Slurm, jobs run at the default frequency, i.e., the maximum frequency, system p1/p2 is considered.

\begin{figure}
 	\centering
 		\includegraphics[width=\columnwidth]{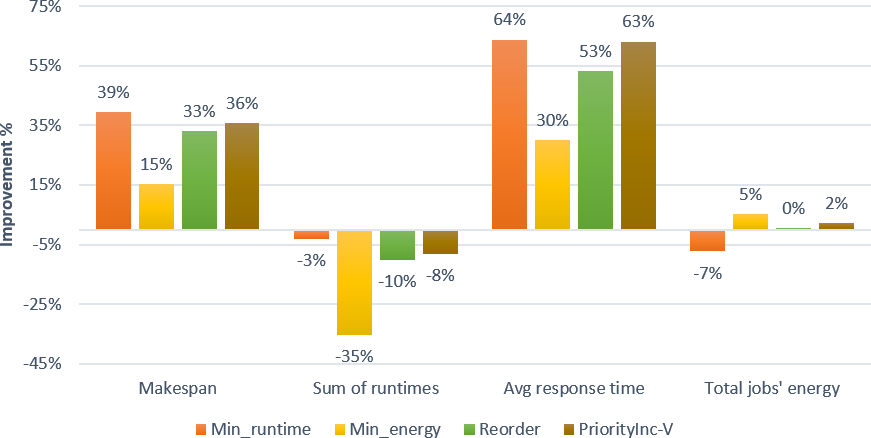}
 	\caption{Savings in terms of makespan, average response time, sum of runtimes and sum of jobs' energy over Base Slurm for EAMC policies, Min\_runtime and Min\_energy.}
 	\label{fig:1part-comp}
\end{figure}

Figure~\ref{fig:1part-comp} shows the improvement in percentage over the Base Slurm for the analyzed metrics.
Applications are equally distributed among the Workload 13\%, and EAMC is configured with \textit{t\_weight}=1.

Analyzing makespan and response time, asking all the available partitions has a considerable impact on all the evaluated scenarios, independently from the job-partition priority, and up to 39\%, and 64\% for Min\_runtime. The poor performance obtained by Base Slurm suggests that the more the systems are interconnected, the better the load can be distributed among clusters. Besides, workload and hardware-aware scheduling can save energy without sacrificing system performance.

Base Slurm can schedule all the jobs in the optimal partition, obtaining the best runtime for each job, but at the cost of the time needed to wait for the favored architecture.
Min\_runtime achieves excellent results for time metrics, at the cost of consumed energy, up 12\% compared to Min\_energy. Min\_energy obtains 5\% energy saving while not scheduling jobs in the optimal performance, but it gives up half of the response time compared to other policies.

Regarding EAMC policies, we can observe that Reorder underperforms slightly compared to other variants because favoring the job's arrival time order leads to fewer possibilities for the scheduler to favor optimal job-partitions. PriorityInc and PriorityInc-V perform similarly to Min\_runtime in terms of makespan and response time while increasing energy savings by 9\%. Increased jobs' runtime, given the lower processors' frequencies, is compensated by the workload-aware scheduling of jobs.

Min\_energy and Min\_runtime only schedule jobs to the first available partition, achieving a low number of running jobs in the optimal partition, as we can observe in Figure~\ref{fig:perc-app}.
EAMC policies, particularly PriorityInc versions, can schedule more jobs in the favored partition, especially in p2, running almost 100\% of STREAM, achieving shorter runtime and higher energy savings.

\subsection{Changing t\_weight parameter for three apps distributions}
As previously commented, seven out of eight applications in our set favor the first partition. In this evaluation, we test the different application distributions: 13\% (uniform distribution), 33\%, and 50\% of jobs benefiting p2. The objective is to evaluate performance for different load levels per optimal partition by changing the number of jobs running the app with id 8, STREAM. We run all the EAMC policies, and we tested two values for t\_weight parameter for Equation~\ref{eq:bestfreq} and~\ref{eq:bestpart}: 1 and 1.5. The latter increases the weight of performance over energy in the EPM by 50\%.

\begin{figure*}
\subfloat[Time metrics for Workload 13\%]{
         \centering
         \includegraphics[width=0.55\textwidth]{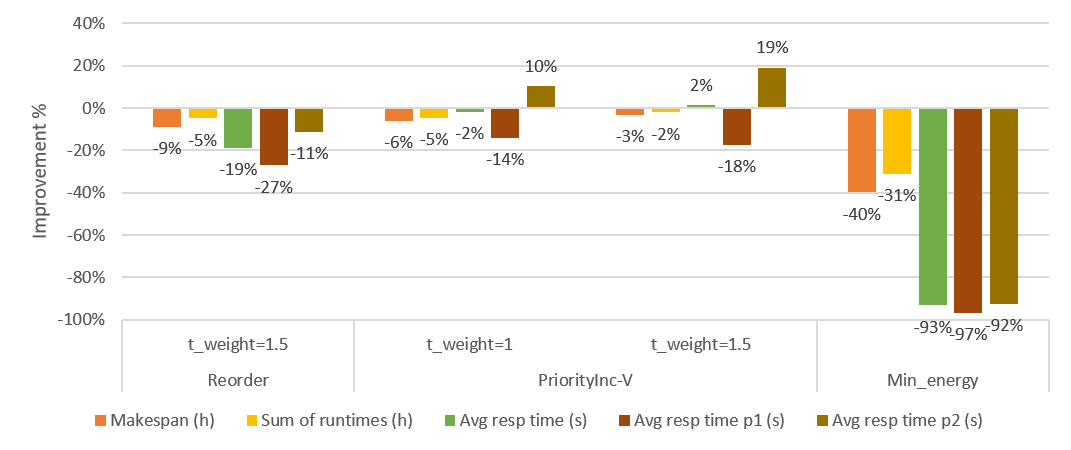}
         \label{fig:metrics-13-time}
    }
     \hfill
     \subfloat[Energy metrics for Workload 13\%]{
         \centering
         \includegraphics[width=0.44\textwidth]{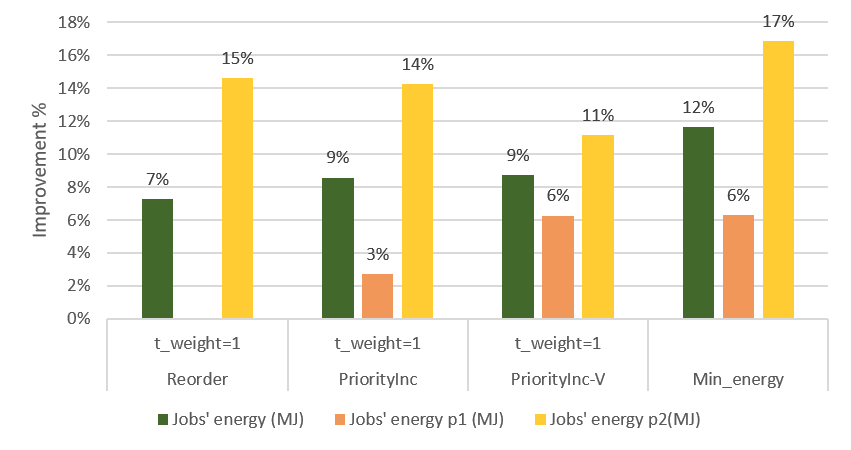}
         \label{fig:metrics-13-energy}
    }\\
    \subfloat[Time metrics for Workload 33\%]{
         \centering
         \includegraphics[width=0.55\textwidth]{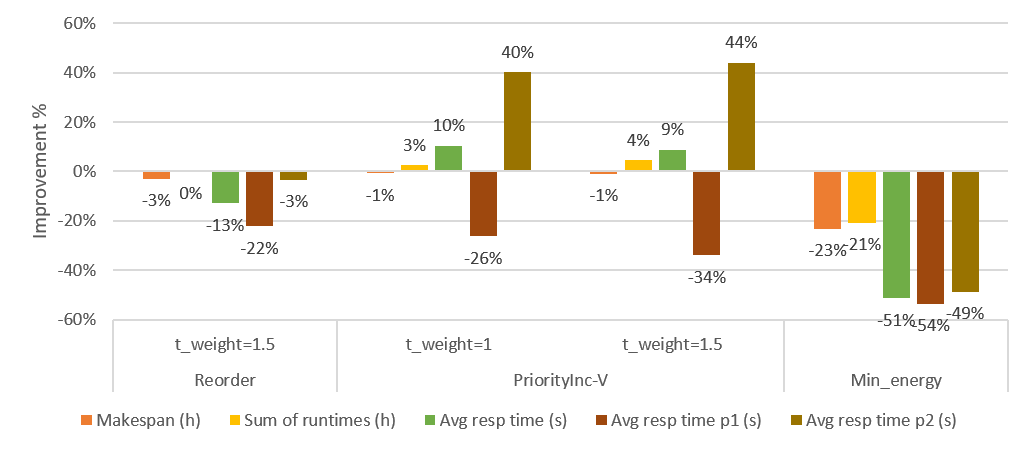}
         \label{fig:metrics-33-time}
    }
     \hfill
     \subfloat[Energy metrics for Workload 33\%]{
         \centering
         \includegraphics[width=0.44\textwidth]{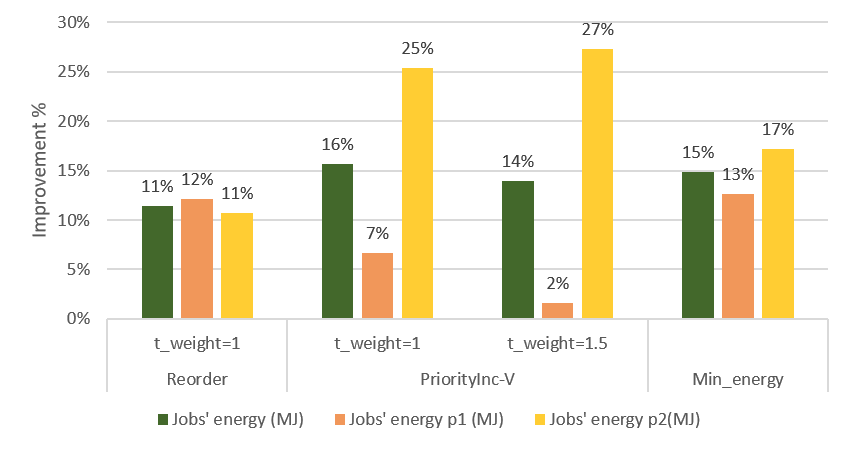}
         \label{fig:metrics-33-energy}
    }\\
    \subfloat[Time metrics for Workload 50\%]{
         \centering
         \includegraphics[width=0.55\textwidth]{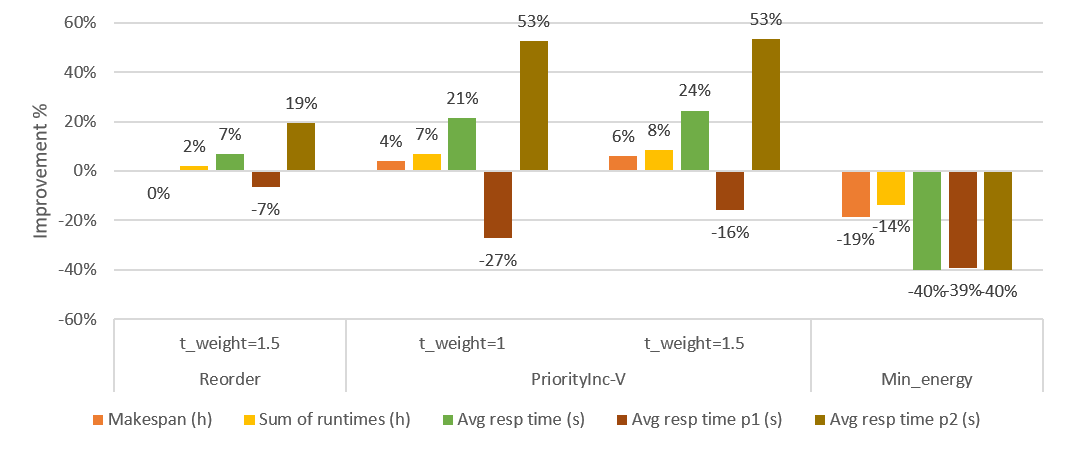}
         \label{fig:metrics-50-time}
    }
     \hfill
     \subfloat[Energy metrics for Workload 50\%]{
         \centering
         \includegraphics[width=0.44\textwidth]{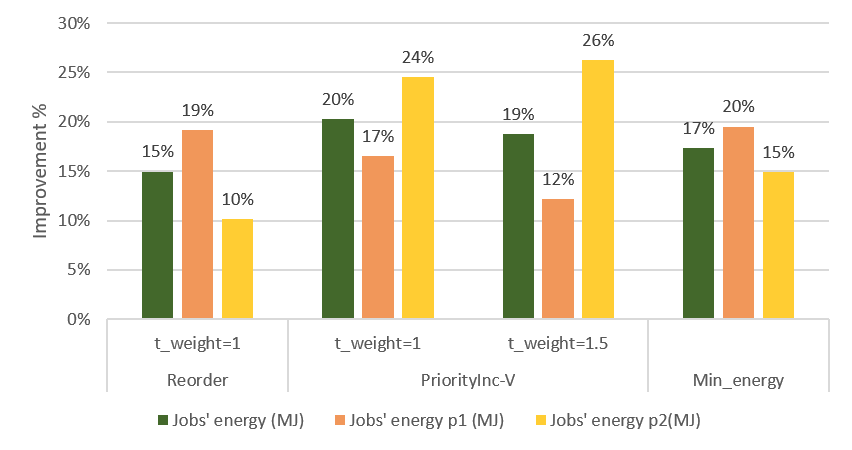}
         \label{fig:metrics-50-energy}
    }
    
    \caption{Main time (left) and energy (right) metrics normalized to Min\_runtime policy and reported as improvement in percentage over it for the first system configuration (p1 and p2). Reported Reorder, PriorityInc-V, and Min\_energy.}
    \label{fig:metrics-13-33-50}
\end{figure*}

This evaluation focuses on the trend between the three workload distributions, presenting more insights on the 33 and 50 scenarios, since we already analyzed Workload 13. Figure~\ref{fig:metrics-13-33-50} reports a summary of time and energy metrics for Workload 13, 33, and 50, normalized to Min\_runtime, and expressed as increase in percentage over it.

\begin{figure}
 	\centering
 		\includegraphics[width=\columnwidth]{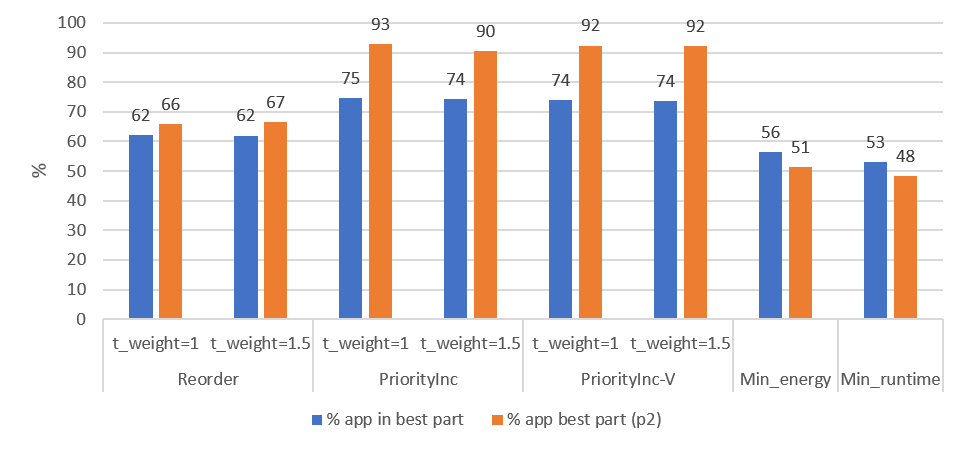}
 	\caption{The average percentage of applications scheduled in the optimal partition for Workloads 13\%, 33\%, and 50\% for the system p1/p2. The average of the two partitions and the partition p2 are presented.}
 	\label{fig:perc-app}
\end{figure}

\begin{figure*}
    \subfloat[Time metrics for Workload 13\%]{
         \centering
         \includegraphics[width=0.55\textwidth]{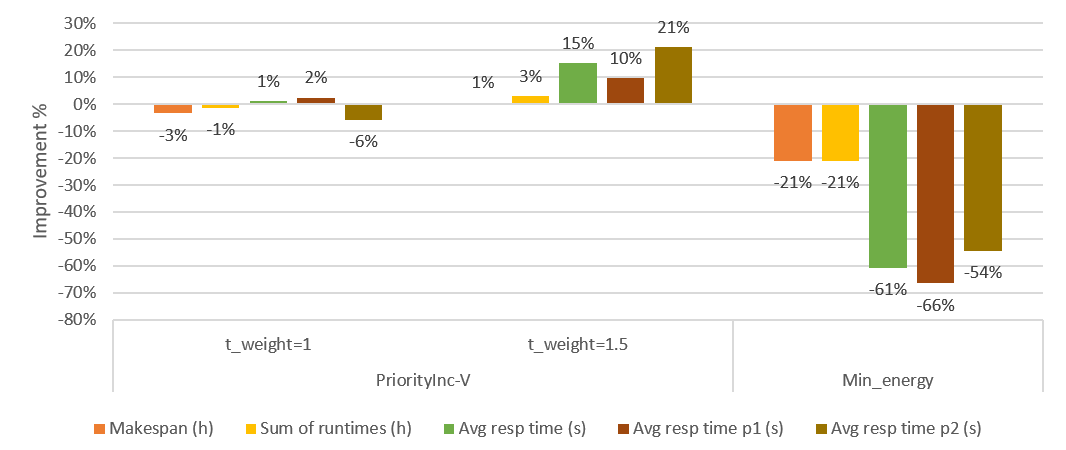}
         \label{fig:metrics-2nd-time}
    }
     \hfill
     \subfloat[Energy metrics for Workload 13\%]{
         \centering
         \includegraphics[width=0.44\textwidth]{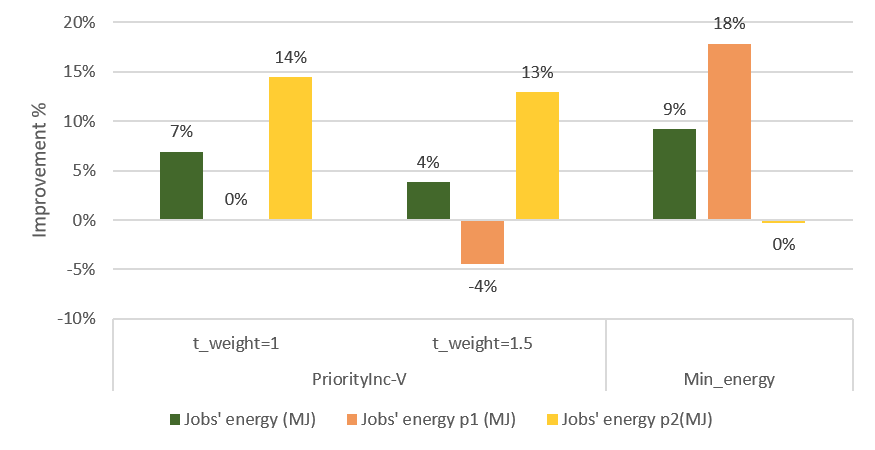}
         \label{fig:metrics-2nd-energy}
    }
    
    \caption{Main time (left) and energy (right) metrics normalized to Min\_runtime policy and reported as improvement in percentage over it for the second system configuration (p1 and p3). Reported PriorityInc-V, and Min\_energy.}
    \label{fig:metrics-2nd}
\end{figure*}

First, moving from distribution 13 to 33 and 50, we notice increasing performance for EAMC when comparing to the other two policies. In those cases, the scheduler can schedule jobs in the optimal partition without sacrificing the response time, as Base Slurm does.

We observe a reduction of response time, by up to 2\%, 10\%, and 25\% respectively for Workloads 13, 33, and 50, when using PriorityInc. Looking at single partitions, we notice that p2 shows the most significant improvement in response time, up to 53\% in PriorityInc-V-1.5. This behavior is compensated by the increased response time in p1, given the lower processor frequency, up to 34\%, but still increasing overall response time thanks to better usage of p2. While the Reorder policy does not reach other EAMC variants in total average response time, it shows more balance in response time between partitions. Gains in response time are accompanied by good results for makespan, ranging from -5\% to 6\% compared to Min\_runtime.

Min\_energy does not perform well in time-related metrics. It increases the response time and the makespan by up to 51\% and 23\% compared to Min\_runtime, and by up to 65\% and 25\% compared to PriorityInc, for Workload 33 and 50, and by 93\% in Workload 13

Min\_runtime energy consumption increases by 17\% and 21\% for Workloads 33 and 50 with respect to Min\_energy. EAMC policies show up to 4\% of improvement over Min\_energy and up to 20\% over Min\_runtime. Energy savings are higher in the second partition, where app8, with almost null frequency scaling, can benefit from a reduced runtime and energy when running over it. PriorityInc-V only slightly improves energy savings over PriorityInc, showing a low number of jobs that perform particularly poorly on a single partition. We identified that app2, ep.D, was affected by the policy variant's threshold out of the eight applications.
Figure~\ref{fig:perc-app} shows the percentage of applications scheduled in the optimal partition. Due to EAMC scheduling, the number of jobs running in the optimal partition reaches 82\%, compared to 55\% in the compared policies, motivating the improvements in the described results. The average frequency for EAMC policies is close to Min\_runtime values concerning p1 and close to Min\_energy in p2. With the increasing number of app8 in Workloads 33 and 50, the average frequency in p2 decreases, as this app's optimal frequency is the lowest.

As a final remark, in Workload 50, Base Slurm average response time and energy consumption improve over Min\_runtime by 19\% and 10\%, at the cost of 7\% higher makespan. This is a specific situation where the load is overall balanced among partitions, but we observed that Base Slurm could not adapt to changes in the load. Besides, while the load is balanced overall, it is not balanced continuously, so EAMC can take advantage of temporary unbalance, further reducing time and energy metrics.

To conclude, PriorityInc-V-1.5 achieves better time-related metrics, with little to no impact on energy consumption. Thanks to the ability to reduce the priority of jobs that have a high impact on performance when running on the secondary partition, it further improves the scheduling. This policy would be our pick for systems on which the performance has primary importance. PriorityInc-1 and PriorityInc-V-1 achieve higher energy saving at a small cost of response time. This pick will be the favorite if energy savings or power costs have the highest importance.

\subsection{Changing t\_weight parameter for system p1/p3}
As a final evaluation, Min\_energy and Min\_runtime to EAMC for the second system configuration. For this evaluation, at maximum frequency, 5 out of 8 applications perform optimally on p1, 3 on p3, while at optimal frequency, all applications perform optimally on p1. Results for Workload 13\% are presented in Figures~\ref{fig:metrics-2nd-time} and Figures~\ref{fig:metrics-2nd-energy}.

For this evaluation, the applications are scheduled first on the optimal partition in the three scenarios, so EAMC-Reorder is not effective like in the precedent evaluation. PriorityInc-V is still able to improve performance in time and energy by reducing the priority of less performant partition. PriorityInc-V with t\_weight=1 improves energy by 7\% on average, while time metrics are comparable. On the other side, PriorityInc-V with t\_weight=1.5 shows 15\% improvements in response time and 4\% of improvement in energy consumption. While Min\_energy achieves 2\% improvement over PriorityInc-V in terms of total energy, it increases the response time by 61\%, a very high overhead.

\section{Conclusions and Future Work}\label{conclusion}
In this article, we presented a new scheduling policy that can estimate the energy and runtime of jobs and use this information to optimize their placement in heterogeneous multi-clusters environments. We used Energy-Aware Runtime and Energy-Aware APIs to connect and monitor applications' behavior. We extended EAR energy model to predict runtime and energy when applying DVFS for multiple hardware architectures.
We used those tools in EAMC-policy, made up of an Energy Prediction Module that predicts energy and runtime for arriving jobs and assigns a different priority to each job-resource combination in heterogeneous environments. Our EAMC Scheduler, based on priority backfill, schedules jobs-resource entities in the same priority assigned by the EPM.
We developed three variants of the policy, and we compared them with the Slurm and policies based on the state of the art.
EAMC-Reorder is able to reduce response time and energy consumption favoring job's arrival order. EAMC-PriorityInc variants are able to optimize the scheduling to further improve makespan, response time and energy savings, up to 6\%, 25\% and 20\% compared to policies minimizing runtime, and up to 26\%, 49\%, and 6\% compared to policies minimizing energy.
As a future work, this policy will be deployed and tested on a production machine, the DEEP-EST prototype, with real workloads, giving the opportunity to generate, collect, and analyze metrics generated by the system as a feedback for the policy.
As a second path, EAMC-policy will be expanded and tested with more heterogeneous systems. To achieve it, a more complex job and hardware modeling need to be performed, not only modeling energy and runtime at different frequencies, but improving the modeling of the behavior of applications for each architecture in terms of performance and amount of resources needed.
Finally, it is interesting to explore how the presented policy behave when integrated with other power-saving techniques, e.g. overprovisioning and powercapping.

\ifCLASSOPTIONcompsoc
  \section*{Acknowledgments}
\else
  \section*{Acknowledgment}
\fi

This work is partially supported by the Spanish Government through Programa Severo Ochoa (SEV-2015-0493), by the Spanish Ministry
of Science and Technology through TIN2015-65316-P project, by the Generalitat de Catalunya (2017-SGR-1414), from the European Union’s Horizon 2020 under grant agreement No. 754304 (DEEP-EST Project), and the BSC-Lenovo collaboration. 

The authors would like to thank Carmen Narvarrete for the kind host at LRZ and her work on the energy interface, and Jordi Aneas Gomez for the applications' data.
\ifCLASSOPTIONcaptionsoff
  \newpage
\fi



\bibliographystyle{IEEEtran}
\bibliography{cas-refs}

%

%
\newpage
\begin{IEEEbiography}[{\includegraphics[width=1in,height=1.25in,clip,keepaspectratio]{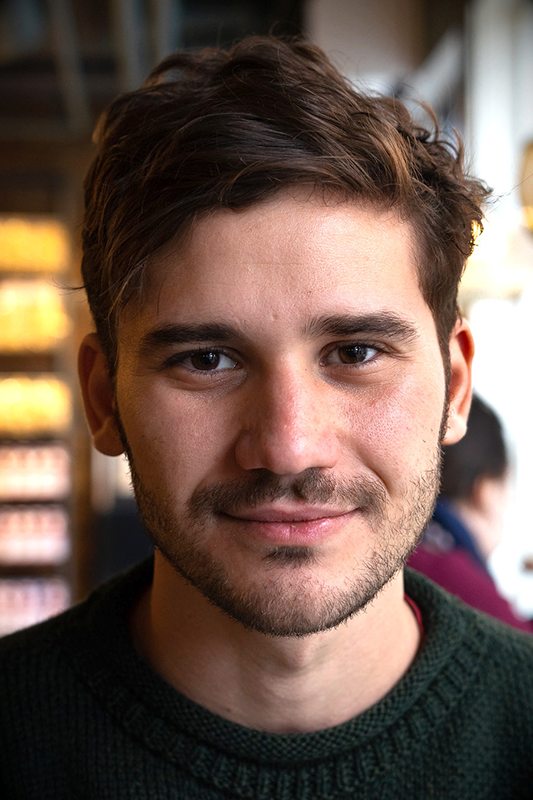}}]{Marco D'Amico} received a BSc and a MSc in computer engineering at Politecnico di Torino, Italy. Internship student as part of the Erasmus+ program at Universitat Politecnica de Catalunya (UPC) in Spain, he is currently enrolled in a PhD program in the same university, and working at Barcelona Supercomputing Center (BSC). His field of research is Job Scheduling for HPC. In particular he researches malleable applications, malleable job scheduling algorithms, runtime metrics collection, and energy-aware job scheduling, with a focus on runtime energy predictions, and energy and power aware job scheduling policies. He works in different European Projects, i.e. the Human Brain Project (HBP) and DEEP-EST Project. He maintains Open Source projects such as the Slurm Simulator. He was awarded with multiple grants, including two internship during his PhD, one month at LRZ and five months at Intel Ireland, where he worked in the hardware field for computer vision.
\end{IEEEbiography}
\begin{IEEEbiography}[{\includegraphics[width=1in,height=1.25in,clip,keepaspectratio]{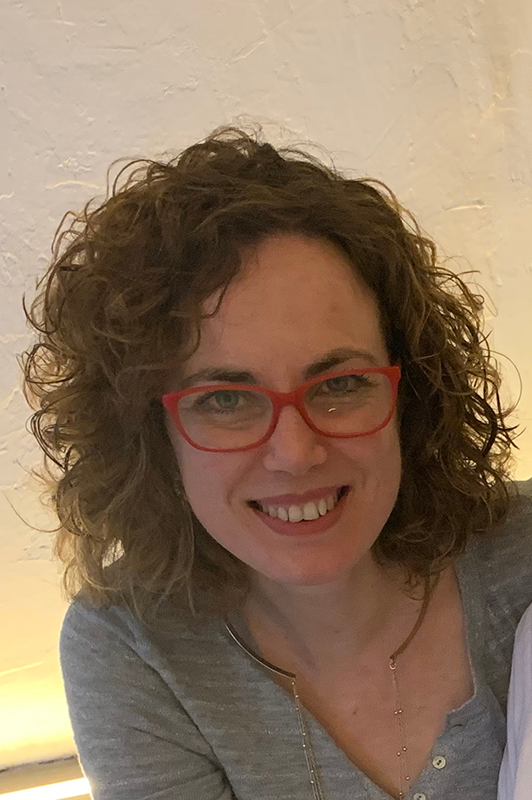}}]{Julita Corbal\'an} received the engineering degree in computer science in 1996 and the PhD degree in computer science in 2002, both from the Technical University of Catalunya (UPC), Spain. Her research interests include processor management of parallel applications, parallel runtimes, and High Performance Computing scheduling policies. She has been the advisor of several PhD in computer science and she has participated in several long-term research projects with other universities and industries, mostly in the framework of the European Union ESPRIT and IST programs. She is currently an associate professor in the Computer Science Department and associate researcher at the Barcelona Supercomputing Center. She is currently involved in two main projects, the BSC-Lenovo collaboration projects, being the leader in the topic of energy aware runtimes and in the DEEP-EST European project designing the new European modular supercomputing architecture.
\end{IEEEbiography}





\end{document}